\documentclass[aps, twocolumn,superscriptaddress]{revtex4-1}

\usepackage{lipsum}
\usepackage{braket}
\usepackage{amsmath}
\usepackage{amssymb}
\usepackage{hyperref}
\usepackage{siunitx}
\usepackage[normalem]{ulem}
\usepackage{graphicx}
\usepackage{colortbl}
\usepackage{xcolor}
\usepackage[normalem]{ulem}

\begin{document}
\title{Predicting quantum dynamical cost landscapes with deep learning}

\author{Mogens Dalgaard}
\affiliation{Department of Physics and Astronomy, Aarhus University, Ny Munkegade 120, 8000 Arhus C, Denmark}

\author{Felix Motzoi}
\email{f.motzoi@fz-juelich.de}
\affiliation{Forschungszentrum J\"ulich, Institute of Quantum Control (PGI-8), D-52425 J\"ulich, Germany}

\author{Jacob Sherson}
\affiliation{Department of Physics and Astronomy, Aarhus University, Ny Munkegade 120, 8000 Arhus C, Denmark}

\date{June 28, 2021}

\begin{abstract}
State-of-the-art quantum algorithms routinely tune dynamically parametrized cost functionals for combinatorics, machine learning, equation-solving, or energy minimization. However, large search complexity often demands many (noisy) quantum measurements, leading to the increasing use of classical probability models to estimate which areas in the cost functional landscape are of highest interest. Introducing deep learning based modelling of the landscape, we demonstrate an order of magnitude increases in accuracy and speed over state-of-the-art Bayesian methods. Moreover, once trained the deep neural network enables the extraction of information at a much faster rate than conventional numerical simulation. This allows for on-the-fly experimental optimizations and detailed classification of complexity and navigability throughout the phase diagram of the landscape. 
 
\end{abstract}

\maketitle
\section{Introduction}

\begin{figure*}
    \centering
    \includegraphics{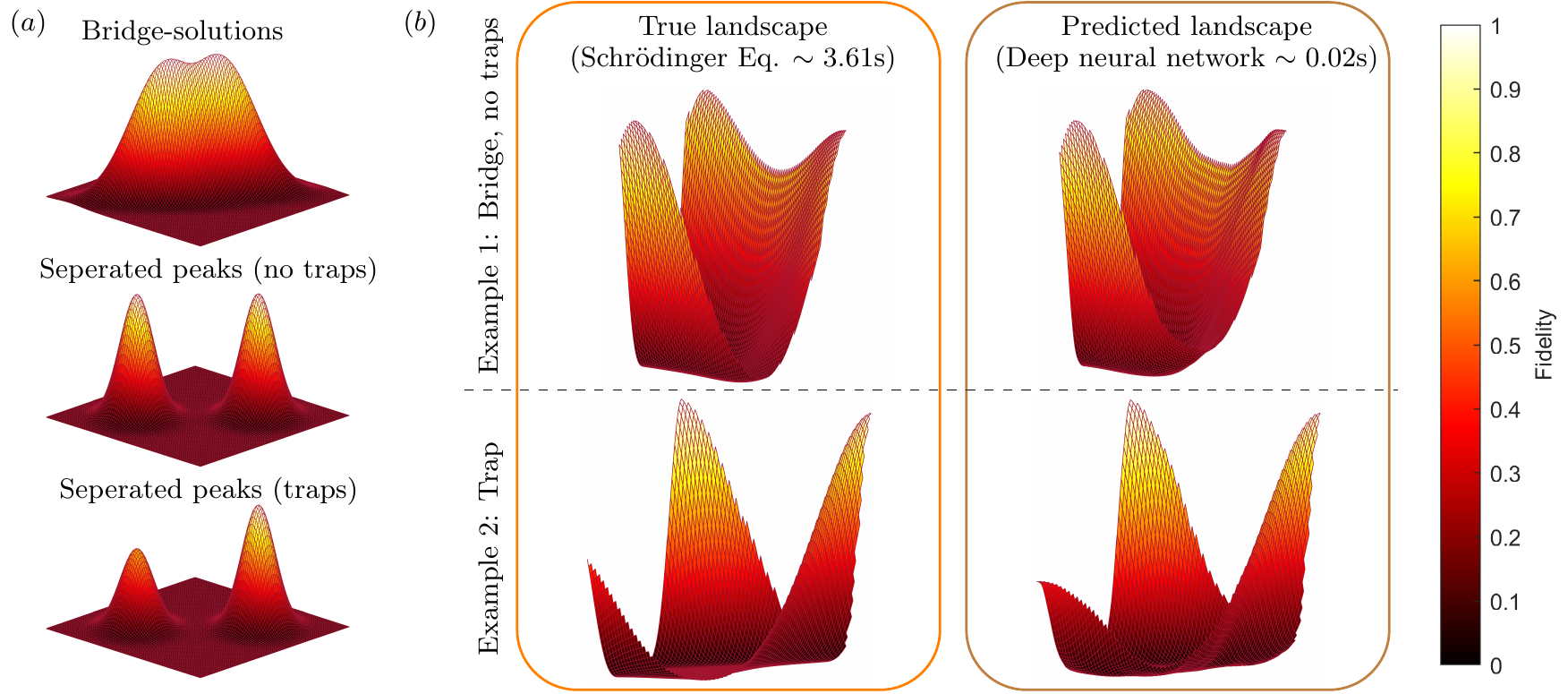}
    \caption{Machine learning the cost functional landscape structure of a spin chain system. The diagrams represent two-dimensional cuts through the control landscape, chosen to include two or three local optima. (a) An illustration of how distinct optima may be connected. They can either be bridged together or constitute separate peaks. In the latter case, they may also be of different heights, with the lower one termed a local trap. (b) The actual fidelity landscape evaluated at all points between three distinct optimized solutions (taken randomly from the validation set) compared to the deep neural network predictions with $g=0$ and $TJ = 4.0$ in Eq.~(\ref{eq:spin_chain_Hamiltonian}). The parentheses denote the computational method and wall time consumption (i.e. the computational time). We select three distinct solutions $v_1$, $v_2$, and $v_3$, which depicts (top) a bridge and (bottom) a trap in the lower left corner.}
    \label{fig:local structure}
\end{figure*}

Many quantum algorithms, notably noisy intermediate scale quantum (NISQ) methods \cite{arute2019quantum,zhong2020quantum,preskill2018quantum, peruzzo2014variational,kandala2017hardware, kokail2019self, farhi2000quantum,omran2019generation, farhi2014quantum, harrigan2021quantum}, encode problem instances onto the parameter space of a quantum device, improving the classical sampling cost. The device acts as a parameter-tunable black box whose output can be optimized via adaptive measurements. Determining the parameter regime to probe typically relies on various heuristics for qualified initial ansätze and subsequent optimal control \cite{glaser2015training,theis2018counteracting, muller2021one} or discrete optimization \cite{maslov2008quantum,motzoi2017linear, yao2020policy, zhu2020adaptive}. Popular heuristic-based algorithms include variational \cite{peruzzo2014variational,kandala2017hardware, kokail2019self}, adiabatic \cite{farhi2000quantum,omran2019generation}, and parametric \cite{farhi2014quantum, harrigan2021quantum} quantum circuits. 

In this context, parameter initialization and optimization may benefit from classical modelling. For example, machine learning can provide a powerful ansatz for the many-body Schrödinger equation given by neural-network encoded quantum states \cite{carleo2017solving, schmitt2020quantum}, be trained to correct for measured quantum error syndromes \cite{liu2019neural}, design improved experiments \cite{goerz2017charting,menke2021automated}, and obtain optimal values for quantum dynamics and circuit parametrizations using reinforcement learning \cite{bukov2018reinforcement,niu2019universal,dalgaard2020global}. These modelling methods focus on parametrizing the behaviour of one or several high-precision candidate solutions.  

In contrast, the black box behaviour of the quantum device (encoding the problem) may itself be learned, or metamodeled, to some helpful degree \cite{wigley2016fast, bentley2018gaussian, kokail2019self, sauvage2020optimal, paulson2020towards,koczor2019quantum}. Here, the modeling task is more challenging since it requires learning the complete (many-to-many) dynamics mapping of the cost functional.
Metamodeling the quantum physical process can reduce ad hoc assumptions about good starting guesses, problem difficulties, and algorithmic hyperparameters. It may also avoid undersampling in areas of interest while oversampling or getting stuck in `barren plateaus' \cite{mcclean2018barren}. Learning a metamodel may also be helpful to transfer knowledge between instances of problems or devices. Perhaps most importantly, offloading the bulk of the modeling cost to the classical co-processor via a metamodel could greatly speed up NISQ devices. 

Discussion of metamodeling is especially relevant to research in the last two decades about cost functional landscapes, particularly for quantum optimal control. Here the most salient open question has been the existence and preponderance of suboptimal solutions (traps) \cite{rabitz2004quantum, pechen2011there, werschnik2007quantum, brif2010control, caneva2011chopped, Larocca_2018,bukov2018reinforcement}, and other measures of problem difficulties \cite{shen2006quantum, hsieh2009topology, nanduri2013exploring,Larocca_2018}, important when selecting suitable optimization algorithms and adjusting their parametrical settings. For instance, discrete optimization methods are well known to exhibit phase transitions between regimes of different difficulty \cite{smith1996locating,xu2000exact,gent1996tsp}, and recent work \cite{bukov2018reinforcement,day2019glassy} has shown that the quantum cost functional landscape of binary-valued (i.e.~bang-bang) control can be mapped onto spin glass physics. This has lead to the proposal of universal behavior of the control landscape under strongly constrained conditions. However, these ideas have yet to be considered in the far more common setting of continuous-value controlled Schrödinger evolution. 

Recent results in metamodeling have focused on Bayesian estimation, using for example Gaussian Process (GP) regression \cite{wigley2016fast,bentley2018gaussian, kokail2019self, sauvage2020optimal, paulson2020towards}, and trigonometric expansion of products of Pauli strings \cite{koczor2019quantum}. The main bottleneck of these approaches is the low precision of Bayesian estimation due to a large computational overhead, in particular when large experimental data sets are needed \cite{paulson2020towards}. In addition, the user-selected covariance function used for fitting may also by restrictive, limiting generalizability and transferability across problem choices. 

In this work, we show that very high precision metamodeling of the complete cost functional landscape can be attained using a deep neural network model. The parametric tunability of the model allows the handling of a much larger data throughput than earlier probabilistic methods, which, as we demonstrate is generally needed to reach error rates at least as low as $10^{-3}$ for the Ising spin chains considered. Even higher throughputs are expected when sampling from quantum devices such as superconducting processors \cite{walter2017rapid}, allowing prediction of the complete landscape to an even higher precision.  We also analyze the underlying structure of the cost functional landscape for the Ising chains considered, and extract key measures that can be used to select suitable optimization algorithms and adjust their parametric setting. In the course of our study, we also identify continuous control phases in the landscapes, where transitions between the phases can be identified as the most challenging regimes, in contrast to earlier studies \cite{tibbetts2012exploring, bukov2018reinforcement}, and generalizing from the binary control case \cite{bukov2018reinforcement,day2019glassy}. 

The paper is organized as follows: in Section \ref{sec:learningspinchaindyn} we demonstrate high precision learning of a dynamical quantum cost functional landscape related to the control of a many-body Ising spin chain. In Section \ref{sec:scalling_speed} we consider the scaling properties of various learning methods and compare to solving the associated equations with matrix exponentiation. In Section \ref{sec:landscape_structure} we analyze the underlying functional landscape structure and in Section \ref{sec:conclusion} we conclude the paper.  

\begin{figure*}
    \centering
    \includegraphics[scale = 0.4]{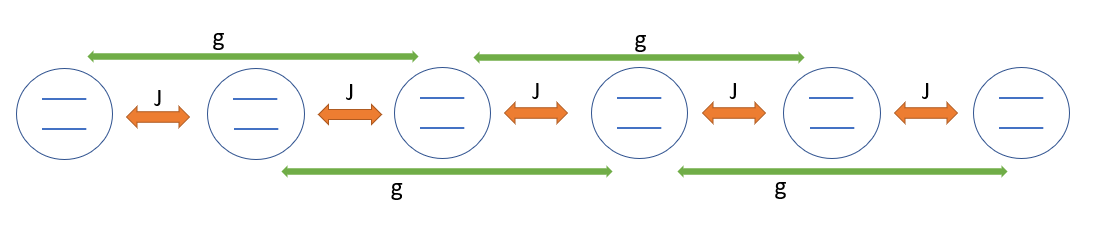}
    \caption{Illustration of a spin chain which consists of a series of coupled two-level systems (spins). Here we model the spin-spin interactions as $ZZ$ terms, where adjacent spins are coupled with strength $J$ and next-nearest spins are coupled with strength $g$. In our simulations, we further assume open boundary conditions.} 
    \label{fig:spin_chain}
\end{figure*}

\section{Learning spin-chain dynamics} \label{sec:learningspinchaindyn}

We study the quantum control cost landscape for a state-to-state transfer of a spin-chain. Here, the objective is to manipulate a quantum system from an initial $\ket{\psi_0}$ to a target state $\ket{\psi_t}$, up to an inconsequential global phase, which is achieved by maximizing the fidelity 
\begin{align}
    F \big[\ket{\psi(T)} \big] = |\braket{\psi_t|\psi(T)}|^2,
    \label{eq:fidelity}
\end{align}
where $\ket{\psi(T)}$ denotes the solution to the Schrödinger equation at final time $T$ starting from the initial state. Here $F = 1$ implies a perfect transfer; and the smallest amount of time where this is possible (at least to a satisfactory degree) is called the quantum speed limit (QSL) \cite{caneva2009optimal}. 

We may induce the state-to-state transfer by controlling the quantum system through externally applied pulses, whose shape can be optimized through quantum control methods in order to reach local optima of the fidelity function Eq.~(\ref{eq:fidelity}). When considering distinct optima, different situation can occur where they may be bridged together as illustrated in Fig.~\ref{fig:local structure}(a, top) or separated peaks as illustrated in Fig.~\ref{fig:local structure}(a, middle). The literature has previously identified the existence of both \cite{Larocca_2018, heck2018remote, larocca2020exploiting}. Note that a bridge could either be a plateau of optima or consist of a set of near-optimal solutions connecting two or more optima. Fig.~\ref{fig:local structure}(a, top) illustrates the latter. Moreover, separated peaks may be of the same height as in Fig.~\ref{fig:local structure}(a, middle) or different heights as in Fig.~\ref{fig:local structure}(a, bottom). In this context, solutions that are not global optima are referred to as traps \cite{rabitz2004quantum,pechen2011there, caneva2011chopped,de2013closer,riviello2015searching,rach2015dressing}.    

In this work we study a one-dimensional spin-chain as illustrated in Fig.~\ref{fig:spin_chain}. The system is governed by the Hamiltonian 
\begin{align}
    H(t) = -J \sum_j \sigma_j^z \sigma_{j+1}^z 
    - g \sum_j \sigma_j^z \sigma_{j+2}^z
    + u(t)\sum_j \sigma_j^x,
    \label{eq:spin_chain_Hamiltonian}
\end{align}
where $J$ denotes the nearest spin-spin interaction, $g$ the next-nearest interaction, and $u(t)$ is proportional to the amplitude of a global transversal magnetic field in the $x$-direction, which we may control. The time evolution we consider occurs via piecewise constant control fields for $N$ equidistant time steps ($\Delta t = T/N$), such that $\ket{\psi(t + \Delta t)} = \exp{(-i H(t) \Delta t)} \ket{\psi(t)}$ where $\hbar \equiv 1$. With this choice, the fidelity functional, Eq.~(\ref{eq:fidelity}), becomes a function of the control amplitudes $F(u_1,u_2,\ldots,u_N)$ and thereby defines the multivariate quantum dynamics landscape \cite{rabitz2004quantum}. We model five spins, with $N = 20$ time steps,  periodic boundary conditions, and limit the time-dependent control $u(t)/J \in [-1,+1]$. The initial and target states are respectively chosen as the two degenerate ground states $\ket{00000}$ and $\ket{11111}$ of Eq.~(\ref{eq:spin_chain_Hamiltonian}) in the absence of control. In this case, the state transfer can only be completed by populating a series of intermediate excited states. In the following, we will investigate the cost functional landscape for two different regimes: without ($g= 0$) and with ($g = J/10$) next-nearest neighbor spin-spin interactions. 

\begin{figure*}
    \centering
    \includegraphics{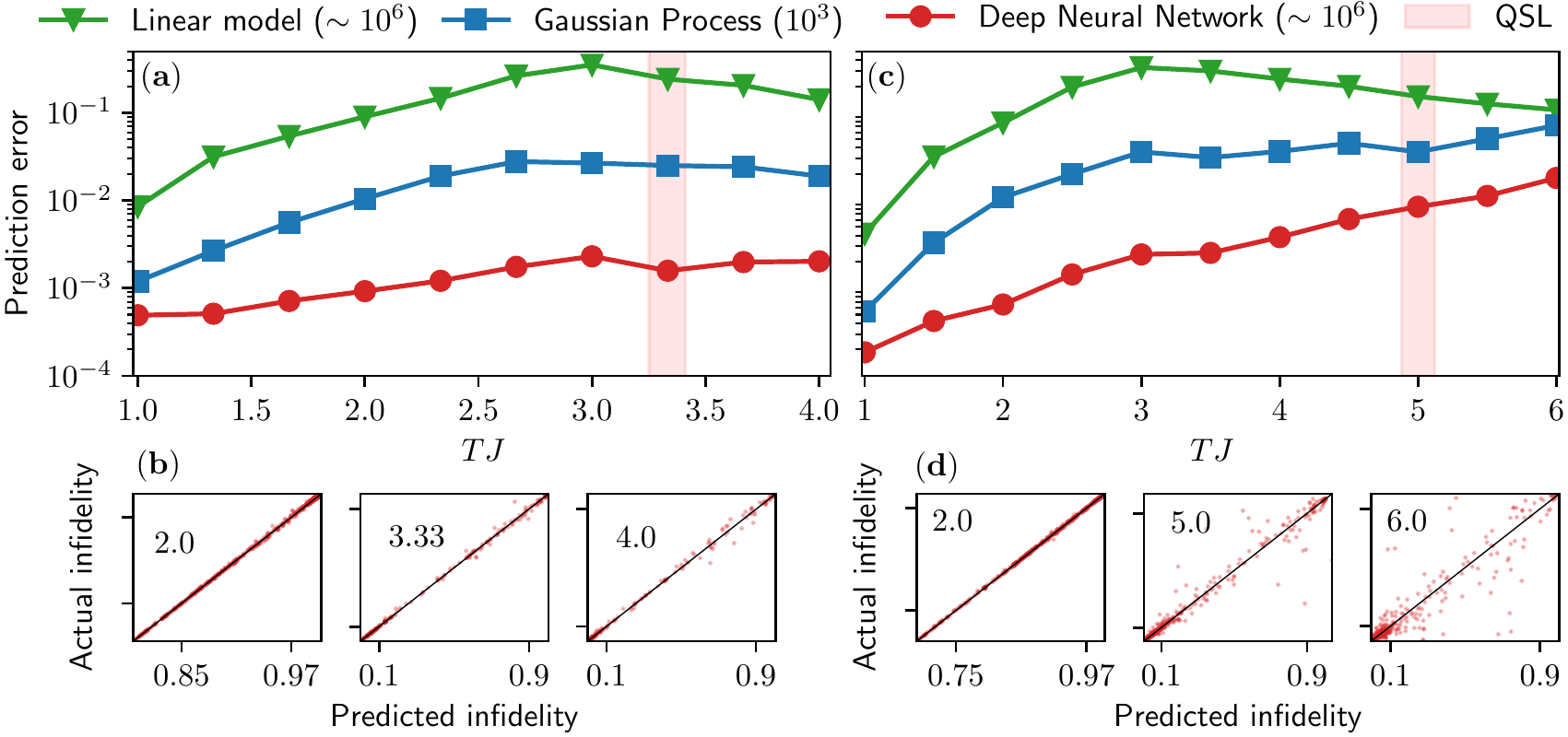}
    \caption{Landscape prediction error as a function of control duration with $N=20$ time steps and $L=5$ spins for (a) nearest neighbor interaction only ($g=0$). Here the parentheses denote the data size, and the shaded region the quantum speed limit (QSL), which we estimate from the collected data. (b) We also plot the predicted versus the actual infidelity at selected durations for the deep neural network for 1000 points drawn at random from the validation set, where the specified numbers denote the total control duration $TJ$. We repeat the analysis for next-nearest neighbor interaction ($g = J/10$) in (c) and (d).}
    \label{fig:spin_chain_learning}
\end{figure*}

The cost functional landscape is not simple enough to fully explore with random sampling alone. Therefore, we utilize the second-order GRAPE optimization algorithm \cite{khaneja2005optimal,motzoi2011optimal,de2011second,dalgaard2020hessian}, which seeks to minimize the infidelity $\mathcal{C} = 1-F$ by performing gradual updates $\mathcal{C} (\mathbf{u}_{n+1}) \leq \mathcal{C} (\mathbf{u}_{n})$ via gradient or Hessian based optimization. We save the infidelity and pulse of each gradual update $(\mathbf{u_n}, \mathcal{C}(\mathbf{u_n}))$ made by GRAPE. This is done for 10,000 randomly drawn pulses at each control duration, which typically results in about one million collected data points. We split the collected data into $80\%$ for training and $20\%$ for validation. The actual structure of the cost functional landscape $\mathcal{C}(\mathbf{u})$ will be discussed further in section \ref{sec:landscape_structure}. 

We seek to learn the fidelity landscape given by Eq.~(\ref{eq:fidelity}) via a deep neural network for which we use a feed forward network with several hidden layers (for implementation details see Appendix \ref{sec:Implementation_details}). The landscape is a very high dimensional (20-dimensional) manifold that cannot easily be visualized. However, we may still depict low-dimensional slices through this landscape. In Fig.~\ref{fig:local structure}(b, top) we compare the predictions of a deep neural network trained on the training set to the true landscape obtained by solving the Schrödinger equation via matrix exponentiation, which is a standard integration technique \cite{blanes2009magnus}. For comparison, we select three distinct, representative optima $v_1, v_2,$ and $v_3$ from the validation set which defines a 2D cut through the landscape, parametrized by
\begin{align}
    v(\alpha,\beta) = v_1 + \alpha(v_2-v_1) + \beta(v_3-v_1).
\end{align}
For both methods we use a $100 \times 100$ equidistant grid with $\alpha,\beta \in [-0.2, 1.2]$, where we plot everything within the physical boundaries ($u_n/J \in [-1,+1]$).

Fig.~\ref{fig:local structure}(b, top) depicts a comparison between the actual solution space and the one predicted by the deep neural network for the same optima $v_1, v_2$, and $v_3$ chosen from the validation set.  From the figure we  see that the deep neural network can very accurately recreate the fidelity landscape, except for small and subtle differences. From the figure we also see the existence of a bridge between two of the solutions. The analysis is repeated for three different solutions in Fig.\ref{fig:local structure}(b, bottom), where we see the existence of a trap, i.e., a solution that is not a global optimum.

Besides deep neural networks, we also assess the predictive power of a different interpolation method, Gaussian processes regression \cite{rasmussen2003gaussian}. Gaussian processes are fast to train and work very well with smaller data sets. In contrast, deep neural networks are known to take longer to train, but can handle much larger amounts of data. We refer the reader to Appendix \ref{sec:Implementation_details} for a technical explanation on how these were implemented. To evaluate the performances, we define the prediction error as the mean over absolute differences $|\mathcal{C}^{\text{pred}}-\mathcal{C}^{\text{actual}}|$ between the predicted infidelity $\mathcal{C}^{\text{pred}}$ and the actual infidelity $\mathcal{C}^{\text{actual}}$ in the validation set. As a comparison baseline, we also use a linear model $\mathcal{C}^{\text{pred}}(\mathbf{u}) = \mathbf{w}^T \mathbf{u} + b$, where the model parameters $(\mathbf{w}, b)$ are found by linear least-square regression.

The collected data constitutes roughly one million pulses and infidelities per control duration. Since Gaussian processes cannot handle such large quantities of data, it is only tested on a subset of 1000 pulses and infidelities. The results with nearest neighbor interactions only ($g = 0$) is depicted in Fig.~\ref{fig:spin_chain_learning}(a). Gaussian processes perform significantly better that the linear model with around an order of magnitude improvement across different control durations. The deep neural network performs even better, with approximately one and two order of magnitude improvements over the Gaussian processes and the linear model respectively. In Fig.~\ref{fig:spin_chain_learning}(a), we show the quantum speed limit (QSL), which we estimate from the collected data to be $T_{\text{QSL}}J \approx 3.33$. To verify that the deep neural network has learned to predict infidelities, we plot in Fig.~\ref{fig:spin_chain_learning}(b) the predicted versus the actual infidelity for 1000 randomly selected pulses in the validation set. For all control durations, the infidelities generally lie on the diagonal line, but with a few points off at $TJ = 4.0$. 

We repeat the analysis with next-nearest neighboring interactions ($g/J=0.1$). This significantly increases the complexity of the learning task and prolongs the quantum speed limit, which is now around $T_{\text{QSL}}J \approx 5.0$. We plot the results in Fig.~\ref{fig:spin_chain_learning}(c), where we see some of the same tendencies as before: the deep neural network performs significantly better than both the linear model and the Gaussian process. However, now the prediction errors for the machine learning algorithms are higher and scale worse with time, although they still achieve overall very accurate predictions. Again, we plot the predicted versus the actual infidelity of the deep neural network for 1000 randomly selected pulses in the validation set at selected control times in Fig.~\ref{fig:spin_chain_learning}(d). Here we observe a region of high (low) predictability at small (large) control durations, which is directly related to the underlying structure of the cost functional landscape. This we investigate further in Section \ref{sec:landscape_structure}, where we relate the landscape structure to Fig.~\ref{fig:spin_chain_learning}.    

\begin{figure*}
    \centering
    \includegraphics{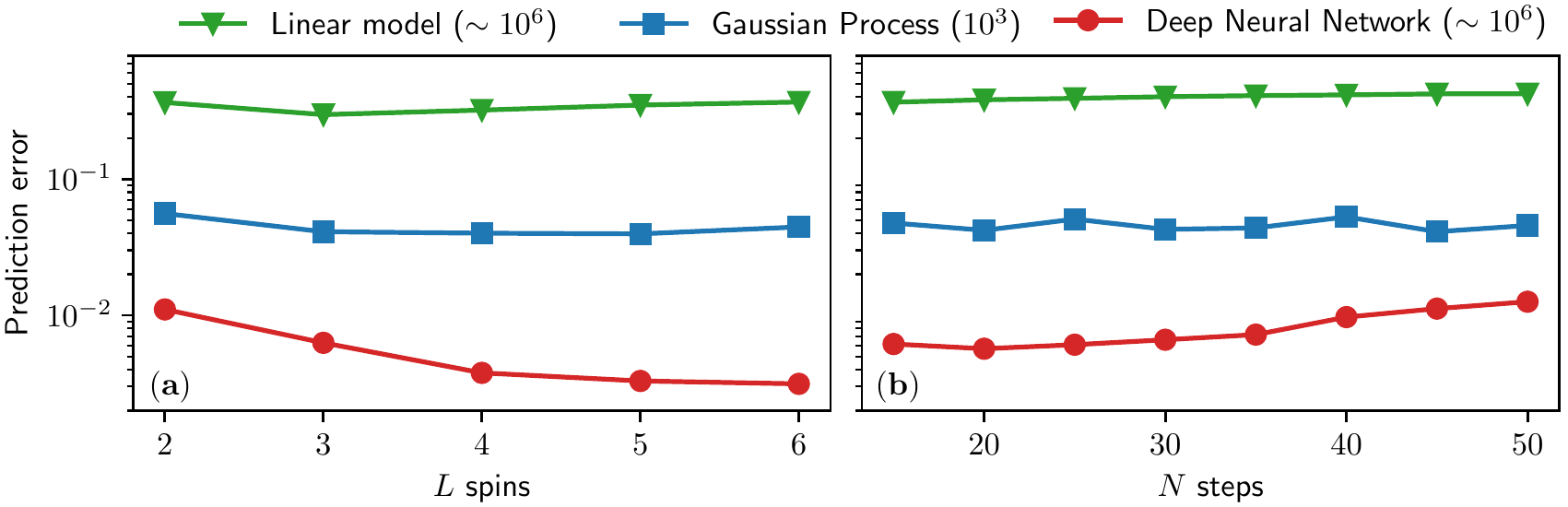}
    \caption{The scaling of the prediction errors when modeling Eq.~\ref{eq:spin_chain_Hamiltonian} with nearest neighbor interaction only (g=0). The prediction error calculated as the average over the difference in predicted versus actual infidelities $|\mathcal{C}^{\text{pred}}-\mathcal{C}^{\text{actual}}|$ from the validation set. ($a$) Scaling with the number of spins $L$ using $N = 20$ time steps at $TJ = 3.33$. Note, the prediction problem for the neural network counterintuitively becomes simpler when more spins are modelled, unlike the Gaussian process. $(b)$ Scaling with the number of time steps $N$ using $L=5$ spins also at $TJ = 3.33$. 
    }
    \label{fig:scaling_errors}
\end{figure*}

\section{Scaling and speed-up} \label{sec:scalling_speed}

Two questions naturally arise with the models presented so far: how does each method scale in performance when the control problem is changed, and are there potential advantages in their use?

To answer these questions, we focus on the example of modelling nearest spin-spin interactions only ($g=0$) at the quantum speed limit ($TJ = 3.33$) for five spins. We start by looking at the performance of the various methods by considering how they scale with the number of spins in the spin chain. This is depicted in Fig.~\ref{fig:scaling_errors}(a) where we have otherwise repeated the procedure from the previous sections. Again, we see the same tendencies as before with the deep neural network significantly outperforming the other methods. However, Fig.~\ref{fig:scaling_errors}(a) also contains a rather counterintuitive result in that the deep neural network, unlike the Gaussian process, performs better with increasing size of the spin-chain, i.e., the control problem seems to become somewhat simpler with increasing chain-size. We will return to this point later in this section. In Fig.~\ref{fig:scaling_errors}(b) we show the scaling with the number of time steps $N$ using $L=5$ spins. Again, we see the same tendencies as earlier when comparing the different methods, but now with a slight increase in the deep neural network prediction error with increasing $N$.  In both cases, we see that increasing the complexity of the dynamics does not significantly affect the ability of the neural network to accurately predict the cost functional landscape.

We now compare the wall time, i.e., the computational time for numerically solving the Schrödinger equation via matrix exponentiation and for evaluating the neural network. Matrix exponentiation scales with the size of the Hilbert space  ($d = 2^L$), whereas the neural network size necessary for encapsulating the dynamics scales with the complexity of the control task. For instance, adding another qubit to the spin chain doubles the size of the Hilbert space, but does not make the control task twice as difficult. As a matter of fact, Fig.~\ref{fig:scaling_errors}(a) points towards the complexity scaling favorably with the number of spins, which we attribute to mean field effects causing the landscape to become simpler with more spins. Note however that the Gaussian process does not see an improvement with increasing $L$, indicating that the model parametrization is key to this effect. 

\begin{figure*}
    \centering
    \includegraphics{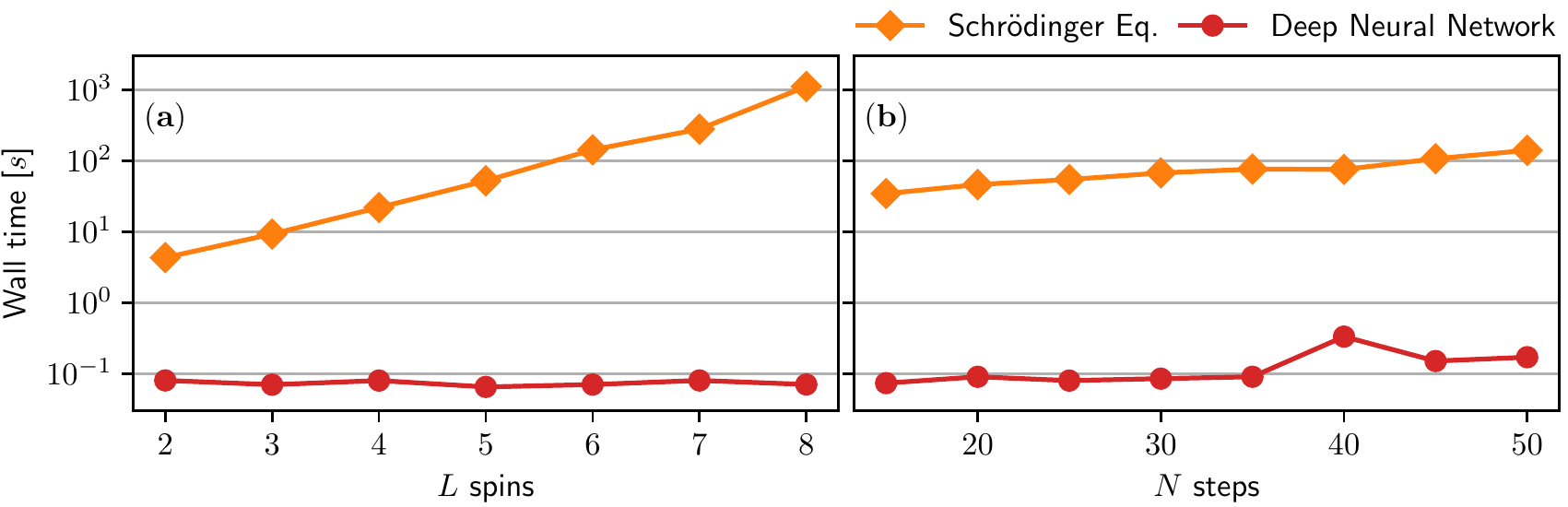}
    \caption{The wall time consumption for numerically solving the Schrödinger equation via matrix exponentiation and evaluating the deep neural network. Here each point corresponds to the evaluation of $1,000$ pulses. ($a$) The scaling with an increasing number of spins $L$. Here we obtain a very favorable scaling with the deep neural network, since its evaluation does not scale with the size of the Hilbert space ($d=2^L$). ($b$) The scaling with an increasing number of time steps $N$, which is also favorable to the deep neural network.  }
    \label{fig:wall_time}
\end{figure*}

In Fig.~\ref{fig:wall_time}(a) we compare the wall time consumption for numerically solving the Schrödinger equation via matrix exponentiation with evaluating the neural network. As expected, the neural network scales independently of the size of the Hilbert space, whereas solving the Schrödinger equation via matrix exponentiation scales very unfavorably with increasing dimensionality, leading to several order of magnitude improvement with the neural network. There exist of course other techniques which allow for handling larger Hilbert spaces \cite{perez2006matrix,de2008density,orus2019tensor}, but these still typically scale with the Hilbert space dimension, due to calculation of the full state dynamics. Hence, using a deep neural network to predict cost functional outcomes could lead to larger computational improvements in applications where many repeated evaluations of the Schrödinger equation are necessary. This must, of course, be compared with the additional time of collecting data and training, for example through experimental cost functional sampling. In Fig.~\ref{fig:wall_time}(b), we also compare the wall time with increasing number of time steps $N$. The width of the neural network scales linearly with $N$ (see Appendix \ref{sec:Implementation_details}) leading to a slight increase in wall time with $N$. However, the neural network still performs much faster than solving the Schrödinger equation with matrix exponentiation. From the figure we also see a fluctuation around $N = 40$, which we believe is due to some numerical instabilities of measuring the wall time.

At the quantum speed limit for $L = 5$ spins and $N = 20$ time steps, collecting the data took around $22$ hours, while the subsequent training of the neural network took around $37$ hours. From this data along with the speed-up depicted in Fig.~\ref{fig:wall_time}(b), we estimate that the two methods become comparable in total computational time when the number of evaluations of the Schrödinger equation exceeds a couple of million, which is the presently considered regime. Moreover, data collection could be accelerated by performing parallel simulations on different processors and both training and prediction of the deep neural network could be accelerated by using modern GPUs rather than CPUs as used in this work.

\begin{figure}
    \centering
    \includegraphics{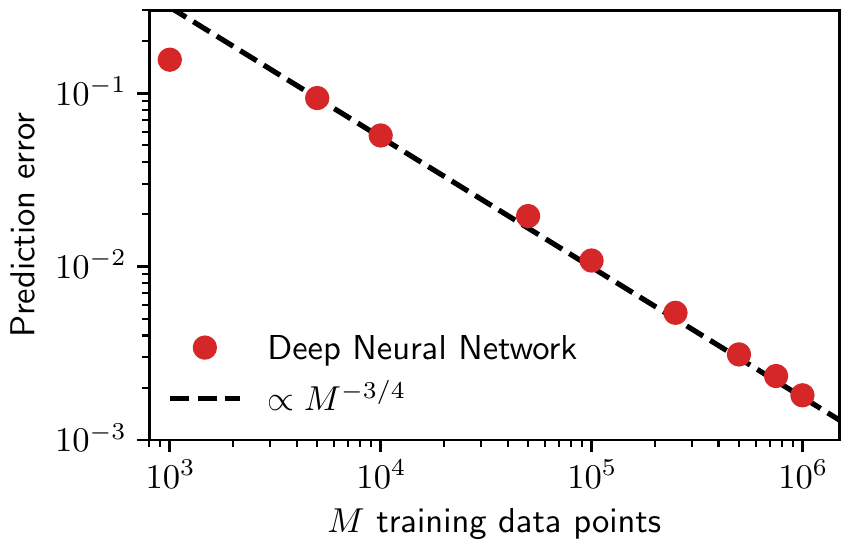}
    \caption{The prediction errors of the deep neural network at different size $M$ of the training data set. The results are consistent with an overall scaling proportional to $M^{-3/4}$, implying that much lower prediction errors are achievable with more training data. }
    \label{fig:training_size}
\end{figure}

In an experimental setting, one can sample the data directly from the quantum device, which for large Hilbert spaces will be much faster than classical sampling from simulations. This can even be done on-the-fly for global optimization, as e.g.~in Ref. \cite{kokail2019self}, by combining with an appropriate optimization algorithm. Once an approximate map of the device is learned, one can also quickly extract general features which may transfer from one device or problem to another. Such general landscape features are discussed in the next section.
In addition to considering the scaling with number of time steps and spins, we also investigate the scaling of the deep neural network with the size of the training set. In Fig.~\ref{fig:training_size} we plot the prediction error of the network at different sizes $M$ of the training set, with $L=5$ spins and $N = 20$ time steps, and at the quantum speed limit $TJ = 3.33$. We make the empirical observation that all of the data points, except the first, lie on a straight line in the log-scale plot, which is consistent with a scaling proportional to $M^{-3/4}$. We depict scaling in the figure with a black dashed line. Hence, increasing the size of the training set by a factor of 10 would reduce the prediction error with a factor of $10^{-3/4} = 0.1778 \sim 1/5$ for this particular problem.
It is especially noteworthy that the error curve plotted in the figure does not saturate within the considered sizes of data, indicating that significantly lower prediction errors would be achievable with access to more data. Although this is outside the computational scope of our computing resources, in an experimental setting, it is possible to obtain larger data sets directly from the quantum device. For example superconducting qubits can have very high data acquisition rates \cite{walter2017rapid}, even scaling favourably with increasing Hilbert space sizes, thus potentially going well beyond the data sizes considered here.

\section{The landscape structure} \label{sec:landscape_structure}

Having obtained a highly accurate mapping of the dynamical quantum fidelity landscape, we turn our attention to extracting some of its most important characteristics.  

\subsection{Landscape properties} \label{sec:landscape_characteristics}

The structure of the cost landscape not only contributes directly to our understanding of the underlying dynamics but is also important from an optimization point of view, since given optimization techniques can have both advantages and disadvantages for different problems. For instance, in Ref.~\cite{dalgaard2020global} machine learning-enhanced exploration outperformed optimization through random seeding on a problem with combinatorially few global optima, while hill-climbing approaches can suffer barren plateaus and vanishing gradients when random sampling is used \cite{mcclean2018barren}.  Furthermore, in the discrete optimization setting it is well known that different problem classes can map onto classical phase transitions exhibiting universal behaviour \cite{smith1996locating,xu2000exact,gent1996tsp}.  This has recently been extended to the discrete control (binary constrained) quantum optimization setting where a connection is made to classical spin glass transitions \cite{bukov2018reinforcement,day2019glassy}.  However, relatively little is known about the continuous control optimization landscape, which is far more common given current experimental capabilities, while also benefiting from greater controllability.  Thus, it is broad significance to extend these notions of problem difficulty to a continuous setting.

\begin{figure}
    \centering
    \includegraphics{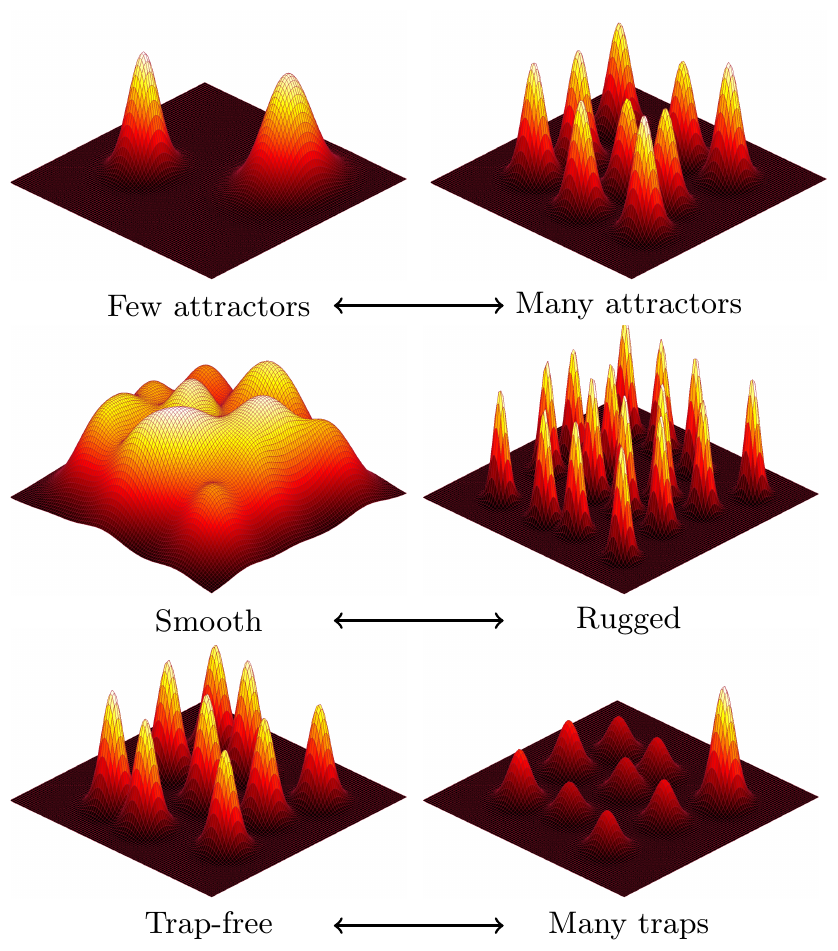}
    \caption{Illustration of opposite regimes of three principle characteristics of the landscape structure. These are: (top) the density of attractors, (middle) the ruggedness, and (bottom) the density of traps. Here the landscape is depicted as a maximization task.
}
    \label{fig:landscape structure}
\end{figure}

Earlier, in Fig.~\ref{fig:local structure}(a) we depicted examples of how the fidelity landscape may look on a local scale. In the following we now seek to investigate the global structure of the landscape. For this purpose, we define a few different measures, which are intended to capture the difficulty of the quantum optimization task as well as the global landscape structure. A schematic of three important characteristics is given in Fig.~\ref{fig:landscape structure} and this provides the basis for our measures, with opposite regimes for each feature being shown.  

The first is a characterization of the distance a local optimizer must travel in the quantum landscape in order to reach an optimimum, which is determined by the \emph{the density of attractors}. An attractor is a local optimum (or connected set of optima), towards which local gradient- or Hessian-based optimization will gradually converge. The landscape will have zero derivatives at the point of an attractor unless if it lies on the boundary of the landscape. 

Secondly, while the density of attractors determines the average distance a local optimizer must traverse before convergence, the speed of convergence to the local attractor may also vary greatly between different landscapes. Generally, local hill-climbing (such as gradient- or Hessian-based) algorithms, may perform larger gradual updates on slowly changing landscapes and therefore converge at a faster rate. Hence, we aim to quantify the \emph{ruggedness} of the terrain, i.e., the rate at which the landscape changes with respect to changes in the control parameters. A landscape that changes slowly we describe as smooth, while we describe a landscape that changes rapidly as rugged. A smooth landscape will also have the additional benefit that functional values near an optimum will not vary greatly, thereby also providing \emph{robustness} to experimental imperfections.

Thirdly, the optimization effort is greatly influenced by the existence of suboptimal local attractors, hence the last characteristic is the \emph{the density of traps}. A trap is an attractor that is not a global optimum, where local gradient- or Hessian-based optimization will still converge towards, i.e.~become trapped. Therefore, finding the global optima would be difficult for a quantum landscape with many traps, often requires a global landscape search strategy. In the opposite regime where there are no traps, local optimization would converge to an optimum with unit probability. For this reason, the existence of traps has attracted much attention in the quantum control literature \cite{rabitz2004quantum,pechen2011there, caneva2011chopped,de2013closer,riviello2015searching,rach2015dressing, jensen2021crowdsourcing}.

\subsection{Landscape measures}

We now relate the landscape characteristics described above and illustrated in Fig.~\ref{fig:landscape structure} to different quantitative measures of control optimization difficulty. The first measure is simply the largest (or best) optimized fidelity $F^*=\max_j F_j$, where $\{F_j\}$ denotes the set of optimized fidelities Eq.~(\ref{eq:fidelity}). This corresponds to the tallest of all the peaks in Fig.~\ref{fig:landscape structure}.

In order to describe the computational cost of local optimization we propose two different measures. The first is the optimization distance, i.e., the distance from the initial randomly drawn pulse to the final optimized pulse using BFGS-based GRAPE

\begin{align}
    D_{\text{attractor}} = \frac{1}{N u_{\max} |\mathcal{D}|}
    \sum_{\mathbf{u}\in \mathcal{D}}
    || 
    \mathbf{u}_{\text{optim}}-\mathbf{u}_{\text{initial}}
    ||
    . \label{eq:attractor_density}
\end{align}
Here $N$ is the number of time steps, $u_{\max} = J$ the amplitude bounds imposed on the pulses, $\mathcal{D}$ denotes the set of saved optimization trajectories with size $|\mathcal{D}|$ (i.e. the number of random seeds optimized), and $|| \bullet ||$ denotes the Euclidean distance. The right hand side of the above equation is simply the normalized Euclidean distance from the initial to the optimized pulse averaged over all iterations. This will be small for a landscape with many attractors and large for a landscape with few attractors. This measure has previously been studied in the literature to assess the straightness of local optimization paths \cite{nanduri2013exploring,Larocca_2018}. 

In addition to this measure, we also consider the fraction of repeated optimized pulses, $1-|\mathcal{D^*}|/|\mathcal{D}|$ as in Ref. \cite{day2019glassy} with $\mathcal{D}^*$ denoting the set of unique optimal pulses. Here we consider two optimized pulses to be different if their normalized Euclidean distance is larger than $10^{-9}$.  

To describe the ruggedness of the control landscape we propose the mean diagonal Hessian element at the optima
\begin{align}
    \rho_{\text{rugged}} = \frac{1}{N|\mathcal{D}|} \sum_{\mathbf{u} \in \mathcal{D}}
    \sum_{j=1}^N \frac{\partial^2 \mathcal{C}}{ \partial  u(t_j)^2}
    \bigg |_{\text{optima}},
 \end{align}
where $\mathcal{C} = 1- F$ denotes the infidelity, with $F$ given by Eq.~(\ref{eq:fidelity}). The Hessian diagonal describes the rate of change of the gradient, which is relatively small for a smooth landscape that changes slowly and in contrast relatively large for a rugged landscape. The eigenvalue decomposition of the Hessian has previously been utilized to analyze quantum control landscapes \cite{shen2006quantum, hsieh2009topology}. For our calculations we use the analytical form of the Hessian found in Ref.~\cite{dalgaard2020hessian}. 

In order to describe the density of traps in the control landscape we use the variance over optimized fidelities, which is zero for a trap-free landscape.

\begin{figure*}
    \centering
    \includegraphics{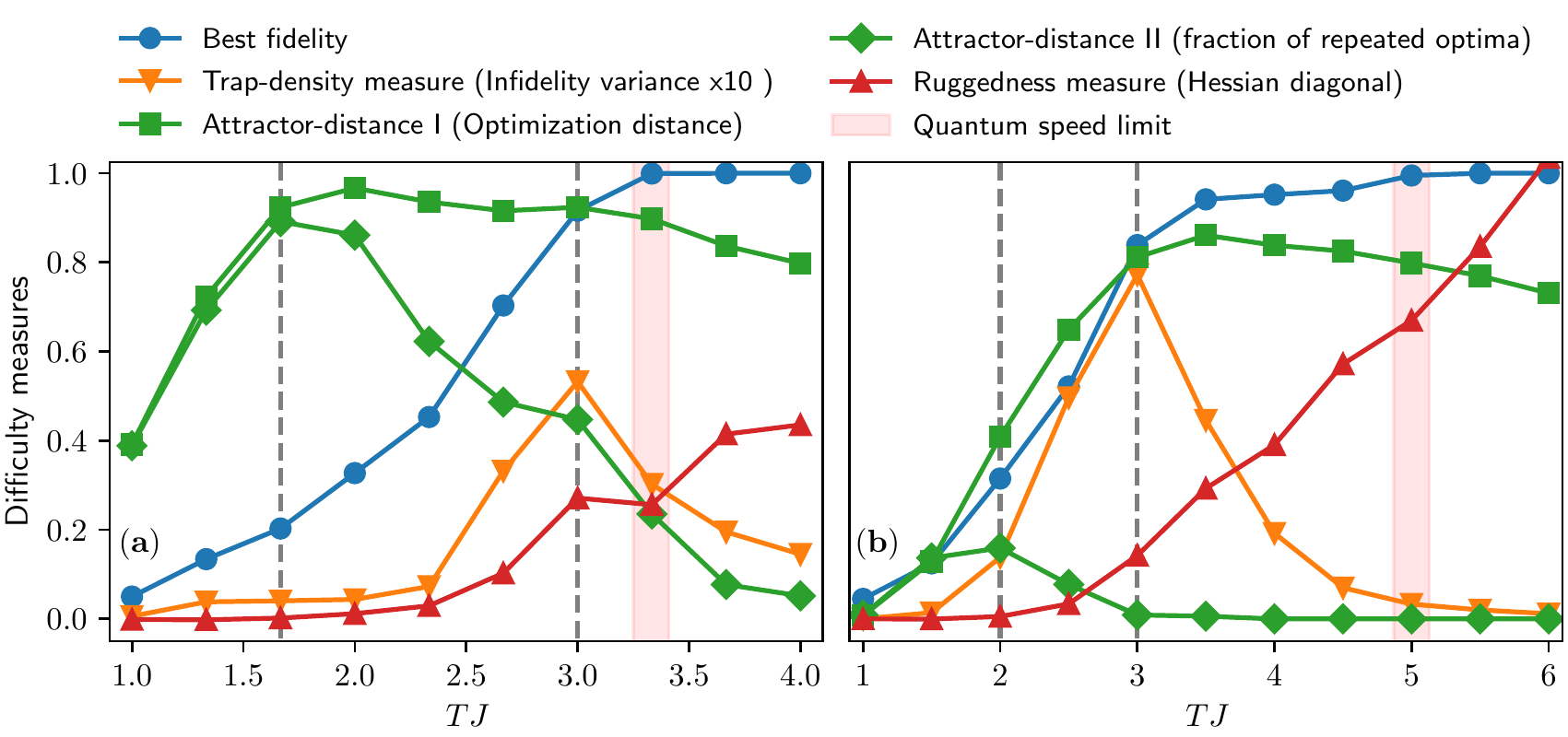}
    \caption{Phases in the control landscape. The best fidelity together with different measures that characterizes the trap-density, attractor-density, and the smoothness for ($a$) nearest spin-spin interaction ($g = 0$) and ($b$) with next nearest spin-spin interaction as well $g = 0.1J$. Here the two vertical dashed lines marks phase transitions i.e., points in the total control duration where the control landscape smoothly changes character, marked by the peak of Attractor-distance II and Trap-density measure (see text). The shaded region marks the quantum speed limit estimated from the data. }
    \label{fig:spin_chain_metric}
\end{figure*}

We will now use the measures given above to determine which characteristics (c.f.~Fig.~\ref{fig:landscape structure}) describe the landscape and its respective phases and relate this to the potential difficulties of numerical or experimental pulse optimization. 

\subsection{Continuous control phases}
We first consider the different phases of the continuous control landscape that emerge as the total control duration is changed. This generalizes on the well-studied cases of discrete optimization \cite{parker2014discrete}, which can be characterized through several measures of optimization-hardness. First, there is computational complexity \cite{bernstein1997quantum}, which is a measure of how the worst-case computational-time effort scales with problem size when finding a global optimum. In contrast, there is also the phase transition language from condensed matter physics, which discusses average properties of (energy) landscapes, and draws conclusions about critical behaviour between different phases or regimes of the landscape \cite{smith1996locating,xu2000exact,gent1996tsp}.  In particular, the latter is more suited for discussion of optimization difficulty since measures of average effort are of the greatest practical relevance (for example the computational complexity does not capture the difficulty in finding a local optimum).

The landscape measures that we use generalize the notion of global vs.~local behaviour, which is a central consideration both for combinatorial optimization and for minimization in energy landscapes. In particular, Ref.~\cite{day2019glassy} identifies two measures related to the Edwards-Anderson order parameter: the first is a local measure that calculates the correlation between optima where improvements are only possible by flipping a single binary control at a time; the second is a more global measure that does the same with two controls at a time.  The authors find that these correspond respectively to phase transitions, from overconstained to correlated, and then to underconstrained optimization. The latter transition also lines up with the quantum speed limit in their system. 

 Because the more general continuous-valued pulses used in our work do not exhibit this discrete bang-bang nature, the measures outlined in the previous section are instead used, indicating both local difficulty (distance and ruggedness) and global difficulty (trapping frequency). These are calculated for different control durations and shown in Fig.~\ref{fig:spin_chain_metric} for both (a) nearest-neighbour interactions ($g=0$) and (b) next-nearest-neighbour interactions ($g = J/10$).

Although our measures are markedly different than the discrete case, we find similar conclusions in terms of over$-,$ critically$-$, and under-constrained dynamics. In both panels, we make the empirical observation that there exist two instances of the total control duration, where the landscapes change behavior, which are the turning points of the density of optima (fraction of repeated solutions) and the density of traps (infidelity variance). This gives us three distinct control regimes, marked with dashed lines, which we can relate to the notion of phases. Note, that the marked phases occur at different times on the two figures.

In essence, we see in Fig.~\ref{fig:spin_chain_metric} that these transitions correspond to two tradeoffs, one corresponding to local effects and the other to global ones. Locally, as the evolution time increases, the number of different dynamical trajectories increases, which can be seen in the increased distance between optima (green squares). Increasing at the same time is the fraction of repeated solutions, as the increased controllability translates into less spatial constraining (green diamonds). These are at odds with each other eventually, and as the distance continues to grow, we see the fraction of repeated solutions abruptly start to drop, indicating the attraction to distinct minima in the landscape. However, these continue to have roughly the same fidelity, as the controls have limited effect at such short durations.
Note also the difference from the discrete case in Ref.~\cite{bukov2018reinforcement}, where overconstraining leads to a convex landscape around a single optimum at short times.  In the present case, the control space (space of possible controls) is significantly larger, leading to a significantly larger set of optimal pulses, even at relatively short control durations.  

The second region sees the appearance of a second tradeoff,  where, as the fraction of distinct optima increases, the globally optimal fidelity also continues to improve (as a result of increased controllability) and so localization of solutions occurs. Therefore, we see the emergence of traps (orange triangles). Abstractly, this can be seen as critical constraining, that is where the search space becomes commensurate with the number of constraints and so satisfying the maximum number of criteria (for maximal fidelity) becomes increasingly difficult. Thus, we see the landscape changing from the situation depicted in Fig.~\ref{fig:local structure}(b) in the top panel increasingly to that of the bottom panel. 

In the third region, the trapping begins to subside as we also see the average distance between optima decreasing from its maximum in Fig.~\ref{fig:spin_chain_metric}. This is the so-called underconstrained phase, where the number of repeated solutions is very small but nonetheless the constraints for high fidelity are still easier to achieve. It is also here that we achieve full controllability, with the quantum speed limit existing in this phase.  In the continuum case, this is also where we expect to see bridge solutions \cite{shen2006quantum, hsieh2009topology}, that is to say, solution spaces described by a continuous symmetry.

Finally, it is worth noting differences with conclusions otherwise drawn in the literature. One key finding is that the quantum speed limit and the phase transition  need not to occur at the same point.  Rather, we see that the maximum trapping frequency region can actually be quite separate from the quantum speed limit, in contrast to the common conjecture~\cite{bukov2018reinforcement,day2019glassy,tibbetts2012exploring}. In particular, this calls into question the generality of the finding in \cite{tibbetts2012exploring} that a superexponential effort is needed near the quantum speed limit, as the trapping frequency may actually be quite small at this control duration.  In the next subsection we examine more closely the question of problem difficulty from an optimization and learning perspective.

\subsection{Optimization difficulty} 
 
The landscape of a difficult optimization task is quantified by having large attractor distance, high trap density, and/or being rugged. In fact, from these considerations, it is plain to see in Fig.~\ref{fig:spin_chain_metric} that the first and second critical durations correspond to where the computational effort of optimization is the largest. Moreover, these are the largest local and global optimization efforts, respectively.

At the first transition, characterized where the relative optimization distance (II) is at a maximum, the other measures of difficulty remain fairly low. In particular, trapping probability is quite low and the landscape is very smooth. We can draw a few conclusions about the computational effort required for different types of algorithms and problems based on the different measures. Because the trapping is quite low, the complexity in terms of the number of controls can be relatively benign. That is, a local search may suffice in the sense that any random control seeding will converge to a near optimal solution with high probability.  This discounts the need for global optimizers in this regime. On the other hand, a low number of repeated solutions can be potentially problematic in terms of the Hilbert space complexity. The large distance between peaks can lead to the emergence of long flat portions of the landscape known as barren plateaus, which may lead to exponential complexity as a function of the number of spins \cite{mcclean2018barren}. In particular, this demands the use of good initial guesses or global machine learning based seeding \cite{dalgaard2020global}. Nonetheless, the smoothness of the landscape indicates that one can improve on basic hill climbers such as simplex search and gradient descent by incorporating information about the curvature of the landscape \cite{dalgaard2020hessian}, with potentially near quadratic improvement in the convergence.  Finally, we note that the distance measure can be used to tune common hyperparameters, such as the size of the initial simplex in Nelder-Mead simplex search \cite{nelder1965simplex}.

The second transition point occurs where the trapping frequency is maximal.  This corresponds in the literature respectively to critical constraining in terms of, e.g., constraint satisfaction problems, and frustration in the minimization of energy for many body physics.  In contrast to the first transition, we see that it is not certain to find a global optimum with local optimization, and therefore random sampling may be insufficient to achieve extremely low infidelity, as needed for certain quantum information tasks. In addition, the Hilbert space complexity may also play a significant role again, with optimization distance (I) near maximal in both panels, and hybrid global-local algorithms \cite{dalgaard2020global} should be implemented with carefully designed stopping criteria for the local optimizer to avoid vanishing gradients. The optimization difficulty may also be linked in this regime to the landscape learning difficulty itself. In Fig.~\ref{fig:spin_chain_learning} we see that the critical points ($TJ\approx3$) are exactly where local maxima occur in the landscape learning error (within a larger trend of increasing difficulty with evolution duration).  In this sense, both training of controls and of fidelity functionals suffer from variance in the peaks and critical constraining. Thus we may argue that the empirically observed difficulty in learning the landscape and minimizing infidelity can act as proxies for each other as difficulty measures.  Interestingly, this does not appear to be the case for the quantum speed limits, where the shaded regions in Fig.~\ref{fig:spin_chain_metric} correspond to minima rather than maxima in Fig.~\ref{fig:spin_chain_learning}.

Finally, we comment on the longstanding debate about the trap frequency in dynamics optimization for typical problems \cite{rabitz2004quantum, pechen2011there, werschnik2007quantum, brif2010control, caneva2011chopped, Larocca_2018,bukov2018reinforcement}.  It is now generally accepted that constraints are indeed commonly leading to trapping in the cost landscape. However, it is remarkable that the `easy' phase of the landscape attributed to \cite{rabitz2004quantum} is not only occurring far to the right of the quantum speed limit but also to its left. Indeed, at shorter times, decoherence is expected to play a much smaller role, and so it may be of more practical relevance. Thus it seems that for this fairly standard problem considered here, the most important region, namely the quantum speed limit, is actually also in the `easy' regime, and therefore not prone to trapping while remaining of highest practical interest. Of course, this does not exclude the evidence that for other problems the quantum speed limit and critical duration may be much closer to each other. Moreover, although strict trapping may not be a problem, the increased ruggedness at these times indicates limited utility for second order methods and the possibility that gradient descent could still suffer from vanishing gradients.


\section{Conclusion} \label{sec:conclusion}

In this work, we have shown that a highly complicated quantum dynamics cost landscape, the fidelity landscape, can be learned with very high precision using deep learning, improving by an order of magnitude over Bayesian estimation based methods, such as Gaussian process regression. Despite being able to correctly predict tens of millions of points to high accuracy, evaluating the neural network remains orders of magnitude faster than simulating the Schrödinger evolution directly. Thus, when combined with direct sampling of experiments, a complete mapping of many typical experimental dynamic protocols should be possible with high throughput. 

This mapping may not only enable on-the-fly global optimization towards true optima in the system, but allows us to faithfully characterize given experimental systems in terms of their phase diagram and difficulty measures.  These measures generalize notions of difficulty from the discrete optimization case, and provide key insights into algorithmic choices, e.g. global vs local, gradient vs.~Hessian, and their hyperparameters, e.g.~distances between optima, smoothness, etc. Combining the developed learning and characterization methods  thus provides valuable tools and insights for both specific instances and wider problem classes of parametric quantum optimization.

\section{Acknowledgements}
The authors would like to thank Carrie Ann Weidner, Marin Bukov and Martino Calzavara for their help in completing this paper.
This work was funded by the Carlsberg Foundation and by the Deutsche Forschungsgemeinschaft (DFG, German Research Foundation) under Germany's Excellence Strategy – Cluster of Excellence Matter and Light for Quantum Computing (ML4Q) EXC 2004/1 – 390534769.  The  numerical  results  presented  in  this  work were  obtained  at  the  Centre for  Scientific  Computing, Aarhus phys.au.dk/forskning/cscaa.

\appendix
\section{Implementation details} \label{sec:Implementation_details}
We used our own implementation of GRAPE, with exact gradients using a L-BFGS-B optimizer from Ref.~\cite{2020SciPy-NMeth}. In the following we briefly comment on Gaussian process regression and deep neural networks.

A Gaussian process regression setup \cite{rasmussen2003gaussian} consists of a previous set of observed data $(X_1, \mathbf{y_1})$, for which we desire to build a model $f$ that allows us to make new predictions $\mathbf{y_2} = f(X_2)$, while still producing the old results $f(X_1) = \mathbf{y_1}$. We treat this via a multivariate normal distribution

\begin{align}
    \begin{bmatrix}
    \mathbf{y}_1\\
    \mathbf{y}_2
    \end{bmatrix}
    \sim
    \mathcal{N}
    \Bigg(
    \begin{bmatrix}
    \boldsymbol{\mu}_1\\
    \boldsymbol{\mu}_2
    \end{bmatrix}
    ,
    \begin{bmatrix}
    \boldsymbol{\Sigma}_{11} & \boldsymbol{\Sigma}_{12}\\
    \boldsymbol{\Sigma}_{21} & \boldsymbol{\Sigma}_{22}
    \end{bmatrix}
    \Bigg),
\end{align}
where $\boldsymbol{\mu}$ and $\boldsymbol{\Sigma}$ denotes the mean and covariances respectively. Where we model the covariances via a kernel function such as a radial basis function $i\neq j$

\begin{align}
    \boldsymbol{\Sigma}_{i,j}
    = k(\mathbf{x_i},\mathbf{x_j}) = \exp \bigg( -\frac{||\mathbf{x_i}-\mathbf{x_j}||^2}{2l^2} \bigg).
\end{align}
Here $l$ denotes a length scale parameter that we along with the mean $\boldsymbol{\mu} = [\boldsymbol{\mu}_1, \boldsymbol{\mu}_2]$ seek to fit in order to maximize the marginal likelihood over the data

\begin{align}
    p(\mathbf{y}|\boldsymbol{\mu}, \boldsymbol{\Sigma})
    = \frac{1}{(2\pi^d |\boldsymbol{\Sigma}|)}
    \exp
    \bigg(
    - \frac{1}{2}
    (\mathbf{y}-\boldsymbol{\mu})^T
    \boldsymbol{\Sigma}^{-1}
    (\mathbf{y}-\boldsymbol{\mu})
    \bigg).
\end{align}
Here $\mathbf{y} = [\mathbf{y}_1, \mathbf{y}_2]$ and $\boldsymbol{\Sigma}$ the matrix containing all covariances. For Gaussian processes we use the implementation from Ref.~\cite{scikit-learn}. We used radial basis functions as kernels and the L-BFGS-B for optimization, with five restarts per fit. 

A neural network consists of layers, where each layer consists of a set of artificial neurons. Each neuron in a given layer receive an input signal $x_j$ for each neuron in the previous layer and based on this calculates an output signal $y$ that is transmitted to each neuron in the next layer and so on. The output signal is calculated via an activation function $y = a(z)$, where $z = \sum_j w_j x_j + b$ where $w$ and $b$ denotes the weights and biases of a given neuron. These constitute the model parameters we seek to fit in order to minimize a given cost function. A deep neural network is a network that has many layers (and perhaps several million parameters) and hence is expensive to evaluate, but with the possible advantage that is may fit highly complicated models. For the deep neural network, we use the implementation from Ref. \cite{abadi2016tensorflow}. The input layer was linear $a(z) = z$, followed by five hidden layers using the ReLu activation function $a(z) = \max(0,z)$, and the width (neurons per layer) of the neural network scaling with the number of time steps as $20N$. The output layer was also linear, and we used biases for all layers. For optimization, we used the Adam optimizer, the loss function was mean squared error, and we used a batch size of 16 with early stopping and a learning rate of $10^{-5}$.

\twocolumngrid
\bibliography{refs}

\end{document}